\documentclass[12pt]{article}

\usepackage{moreverb}
\usepackage{url}
\usepackage{natbib}
\usepackage{subfigure}
\usepackage{bbm}
\usepackage{comment}

\usepackage{amsfonts,graphicx}
\usepackage[margin=2.5cm]{geometry}
\usepackage[font={small}]{caption}

\newcommand\BibTeX{{\rmfamily B\kern-.05em \textsc{i\kern-.025em b}\kern-.08em
T\kern-.1667em\lower.7ex\hbox{E}\kern-.125emX}}

\newcommand{\real}{\mathbb{R}}
\newcommand{\dd}{\textup{d}}
\begin{document}

\title{\bfseries Calibration diagnostics for point process models via the probability integral transform}

\author{T.~L.~Thorarinsdottir\footnote{Norwegian Computing Center, P.O. Box 114 Blindern, 0314 Oslo, Norway}}
%\author{T.~L.~Thorarinsdottir\affil{a,}\corrauth}
%\address{\affilnum{a}Norwegian Computing Center, P.O. Box 114 Blindern, 0314 Oslo, Norway}
%\corremail{thordis@nr.no}

%\received{00 Month 2013} \accepted{00 Month 2013}

\maketitle

\begin{abstract}
\noindent
We propose the use of the probability integral transform (PIT) for model validation in point process models.  The simple PIT diagnostics assess the calibration of the model and can detect inconsistencies in both the intensity and the interaction structure.  For the Poisson model, the PIT diagnostics can be calculated explicitly.  Generally, the calibration may be assessed empirically based on random draws from the model and the method applies to processes of any dimension.  
\end{abstract}

%\keywords{calibration; model validation; probability integral transform; point process; verification rank histogram}

\section{Introduction}

Point process methodology is applied in diverse scientific fields to model and predict earthquakes, wildfires, disease occurrences, telecommunications, plant and cellular systems, and animal colonies, to name but a few examples. See, for instance, \cite{Andrews&2011}, \cite{Eberhard&2012}, \cite{Edelman2012}, \cite{Klaver&2012}, \cite{Mohler&2011}, \cite{PourtaheriVahidi-Asl2011} and \cite{Waller&2011} for a non-exhaustive list of some recent applications of spatial and space-time point processes. With such variety in the applied fields that use these models as well as the wide range of available models, it is important to have general and -- preferably -- easily applicable methods to assess goodness-of-fit and predictive performance.
 
Most current model validation methods are based on residual analysis.  \cite{Baddeley&2005} apply the Papangelou conditional intensity of a spatial process to define a pixel-based residual diagnostic framework similar to that commonly applied to Poisson log-linear regression, where the observed number of points within each pixel is compared to the estimated number. Further theoretical properties of the residuals, originally proposed by \cite{StoyanGrabarnik1991}, are discussed in \cite{Baddeley&2008} and an extension to space-time models is provided by \cite{Zhuang2006}.  Based on similar principles, \cite{WongSchoenberg2009} compare the pixel-wise log-likelihoods of two competing models. 

Alternatively, the point pattern may be transformed by rescaling, thinning, superposition or superthinning according to the estimated model.  Under an appropriate model, the new transformed pattern is a homogeneous Poisson process, see \cite{Clements&2011} and references therein.  Summary statistics such as the $K$-function \citep{Ripley1977} may then be used to assess the homogeneity of the transformed pattern.  \cite{Baddeley&2011} propose using the $K$-function directly for score tests and residual diagnostics.  Further goodness-of-fit tests and diagnostic tools for Poisson and related processes are, for instance, discussed in \cite{Lawson1993}, \cite{Guan2008}, \cite{Baddeley&2012} and \cite{Baddeley&2012b}. 

We propose a simple, yet effective pixel-wise model validation for point process models based on the probability integral transform (PIT).  In \cite{Dawid1984}, the PIT is proposed as a tool for assessing model calibration which essentially requires the observation to be indistinguishable from a random draw from the model.  That is, if the continuous distribution $F$ is the true distribution of the random variable $Z$ it follows for the PIT value $F(Z)$ that  $F(Z) \sim \mathcal{U}([0,1])$.  For a large number of observations, the calibration may be checked empirically by plotting the histogram of the PIT values and checking for uniformity, see \cite{Gneiting&2007}.  While calibration diagnostics of this type are usually applied to assess the calibration of predictive distributions, the same framework may also be used to assess goodness-of-fit.  In the point process setting, this corresponds to calculating the PIT value $F(Z)$ for each pixel, where $Z$ is the observed number of points in the pixel and $F$ is the estimated distribution of the number of points in that pixel.  

Our method differs from many similar pixel-based validation methods in that we compare the observed number of points to its estimated distribution rather than the expected value.  Related calibration diagnostics for spatial point processes have recently been proposed by \cite{Wong&2012} where the distribution of the observed Voronoi tessellations is compared to the theoretical distribution of tessellations under the model.  

The remainder of the paper is organized as follows.  In Section 2, we introduce the PIT for count data and discuss estimation procedures when the distribution $F$ is not explicitly known.    In Section 3, we apply our calibration diagnostics framework to three examples of simulated data previously studied in \cite{Baddeley&2005} and \cite{Baddeley&2011} under a variety of models. The paper then closes with a short discussion in Section 4.   

\section{Calibration of point count estimates}

For the clarity of exposition, we focus on spatial point processes both in our discussion below and in the examples in the following section. However, the general framework applies to point process models in any dimension, in particular to temporal and space-time models. 

\subsection{Poisson processes}

Let $\Phi$ be a spatial Poisson point process with intensity function $\lambda : \real^2 \rightarrow \real_+$ and assume that we have observed the realized point pattern $\varphi$ on a finite observation window $W \subset \real^2$ which may be divided in disjoint pixels $A_1,\ldots,A_S$.  We denote by $z_s$ the observed number of points in pixel $A_s$.  The expected number of points in each  $A_s$ follows a Poisson distribution $G_s$ with parameter  
\begin{equation}\label{eq:poisson parameter}
\Lambda(A_s) = \int_{A_s} \lambda(u) d u.
\end{equation}  

Under a Poisson model $M$, denote by $F_1, \ldots, F_S$ the estimated distributions for the number of points in $A_1, \ldots, A_S$.  To validate $M$, we assess its calibration, that is, the statistical consistency between $F_1, \ldots, F_S$ and $G_1, \ldots, G_S$.  As discussed in \cite{Gneiting&2007}, consistency holds asymptotically in $S$ for continuous distributions if the sequence $(F_s)_{s=1,2,\ldots}$ is probabilistically calibrated relative to the sequence $(G_s)_{s=1,2,\ldots}$.  That is, if 
\begin{equation}
\lim_{S \rightarrow \infty} \frac{1}{S} \sum_{s=1}^S G_s \circ F_s^{-1}(p) \rightarrow p
\end{equation}
holds almost surely for all $p \in (0,1)$.  For finite samples, the calibration is assessed empirically by comparing the distribution of $F_1(z_1),\ldots,F_S(z_S)$ to that of a standard uniform distribution.  For count data, \cite{Czado&2009} advocate the use of the randomized PIT,
\begin{equation}\label{eq:PIT}
F(Z-1) + V [ F(Z) - F(Z-1)],
\end{equation}
where $V \sim \mathcal{U}([0,1])$ and, by definition, $F(-1) = 0$.  

We empirically assess the uniformity of the randomized PIT values in two ways.  A spatial map of the PIT values across the observation window $W$ may reveal divergences from spatial homogeneity while a histogram of the values may show overall divergences from uniformity.  Note that a uniform histogram is necessary but not sufficient for a calibrated model.  Similarly, these approaches are appropriate for small sample sizes and notable departures from uniformity while small deviations from uniformity require larger sample sizes, see  \cite{Gneiting&2007} and references therein as well as the examples below.  

\subsection{General processes}

When the distribution $F_s$ cannot be calculated explicitly, we replace it by the empirical distribution $\hat{F}_s$ obtained from $K$ samples from the model $M$.  This corresponds to calculating the rank of $z_s$ in $\hat{z}^1_s, \ldots \hat{z}^K_s$, where $\hat{z}^k_s$ denotes the number of points that fall in pixel $s$ in sample $k$ from $M$.  As in the randomized PIT in (\ref{eq:PIT}), we solve ties at random.  This approach to assessing calibration has become a standard procedure, for instance, in weather forecasting where complex numerical models are run multiple times to yield a forecast ensemble that approximates the predictive distribution.  In this setting, the resulting histograms are called verification rank histograms or Talagrand diagrams \citep{Anderson1996, HamillColucci1997} and they may be interpreted in the same manner as the PIT histograms. 

In this form, our calibration diagnostic is a generalization of the Number-test (N-test) proposed by \cite{Schorlemmer&2007}.  The N-test examines the fraction of simulations from $M$ that contain fewer points than the observed pattern $\varphi$ over the entire observation window $W$.  The test statistic is given by
\begin{equation}
\delta = \frac{1}{K} \sum_{k=1}^K \mathbbm{1} \{ n(\psi^k) < n(\varphi) \},
\end{equation}  
where $\mathbbm{1}$ is the indicator function and $n(\psi^k)$ denotes the number of points in the simulated pattern $\psi^k$, and $M$ is considered inconsistent with the data if $\delta$ is close to $0$ or $1$.  As noted by \cite{Clements&2011}, the N-test can provide an overall assessment of $M$ while it cannot indicate where the model may be fitting poorly.  
 
Under the Poisson model, the point counts in the disjoint sets $A_1, \ldots, A_S$ are independent and thus also the randomized PIT values in (\ref{eq:PIT}).  We may therefore employ formal tests of uniformity, see e.g. \cite{Anderson1996} and \cite{CorradiSwanson2006}.  However, for point process models with complex interaction structure the interdependencies complicate the use of formal tests.  A thorough discussion of the potential fallacies regarding the use of verification rank histograms is given in \cite{Hamill2001}.  

\section{Examples}

We test the PIT calibration diagnostics on three simulated examples of spatial point processes previously studied in \cite{Baddeley&2005} and \cite{Baddeley&2011}.  All parameter estimation and pattern simulation is performed using the {\tt R} package {\tt spatstat} \citep{R2013, BaddeleyTurner2005}.  The examples present the three main types of point patterns -- random, repulsive, and clustered -- and include both homogeneous and inhomogeneous cases. 

\subsection{Inhomogeneous process with repulsion}

\begin{figure}[!hbpt]
\centering
\includegraphics[width=\textwidth]{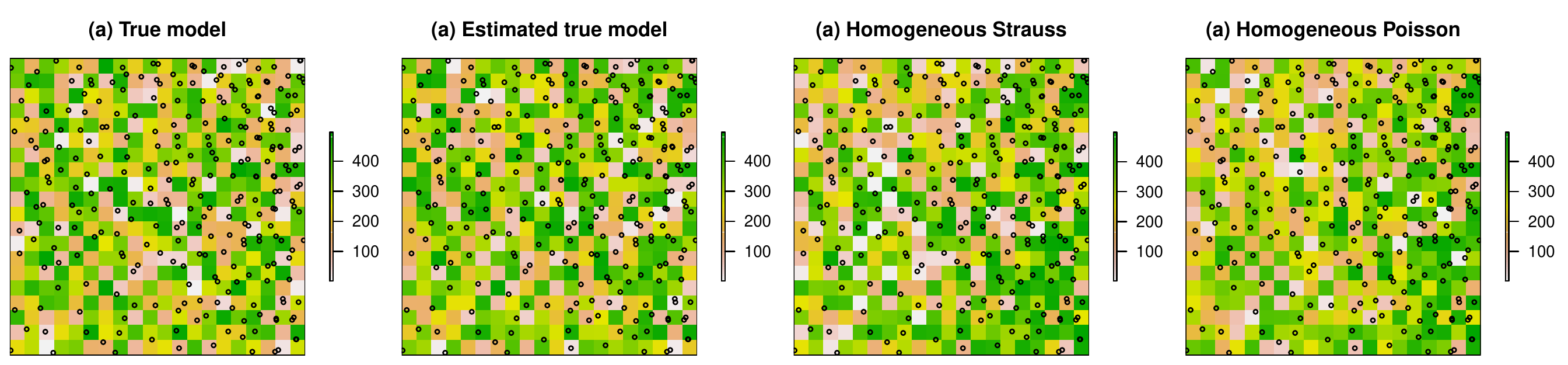}
\caption{Spatial maps of the observed ranks of a pattern drawn from the Strauss model in (\ref{eq:Strauss}) under (a) the true model, (b) the estimated correctly specified model, (c) a homogeneous Strauss model and (d) a homogeneous Poisson model with the parameter estimates given in the main text.  The ranking is based on $499$ samples from each model and the ranks thus range from $1$ to $500$.  The point pattern is indicated with small circles.}
\label{fig:Strauss}
\end{figure}

Our repulsion example is an inhomogeneous Strauss process \citep{Strauss1975} with density 
\begin{equation}\label{eq:Strauss}
p_{\theta}(\varphi) = \mathcal{Z}(\theta) \gamma^{s_{\theta}(\varphi)} \prod_{i=1}^n b_{\theta}(\xi^i), 
\end{equation}
where $\gamma$ denotes the interaction strength, the interaction function is $s_{\theta}(\varphi) = \sum_{i \neq j}\mathbbm{1}\{ \| \xi^i - \xi^j \| \leq r\}$ with an interaction radius $r$, $b_{\theta}$ is an inhomogeneous activity function, and the unknown normalizing constant is denoted by $\mathcal{Z}$.  Specifically, we set $\gamma = 0.1$, $r= 0.05$ and define $b_{\theta}(u) = \theta_1 \exp(\theta_2u_1 + \theta_3 u_2 + \theta_4 u_1^2) = 200 \exp (2u_1 + 2u_2 + 3u_1^2)$. 
The interaction radius is assumed known while the remaining parameters $\theta = (\gamma, \theta_1, \theta_2, \theta_3, \theta_4)$ are estimated by maximizing the pseudolikelihood \citep{Besag1975, BaddeleyTurner2000}.  For the point pattern in Figure~\ref{fig:Strauss}, we obtain $\hat{\theta} = (0.24, 179, 1.53, 1.89, 1.34)$.  We compare the calibration of the true and the estimated correctly specified model to that of a homogeneous Strauss process with $(\hat{\gamma}, \hat{\theta}_1) = (1099, 0.34)$ and a homogeneous Poisson process with estimated intensity $\hat{\lambda} = 271$,  the number of points in the pattern. To estimate the PITs, we simulate $499$ patterns under each model and calculate the rank of the number of observed points in each of $400$ equally sized pixels with the number of observed points in each pixel ranging from $0$ to $3$. 

\begin{figure}[t]
\centering
\includegraphics[width=\textwidth]{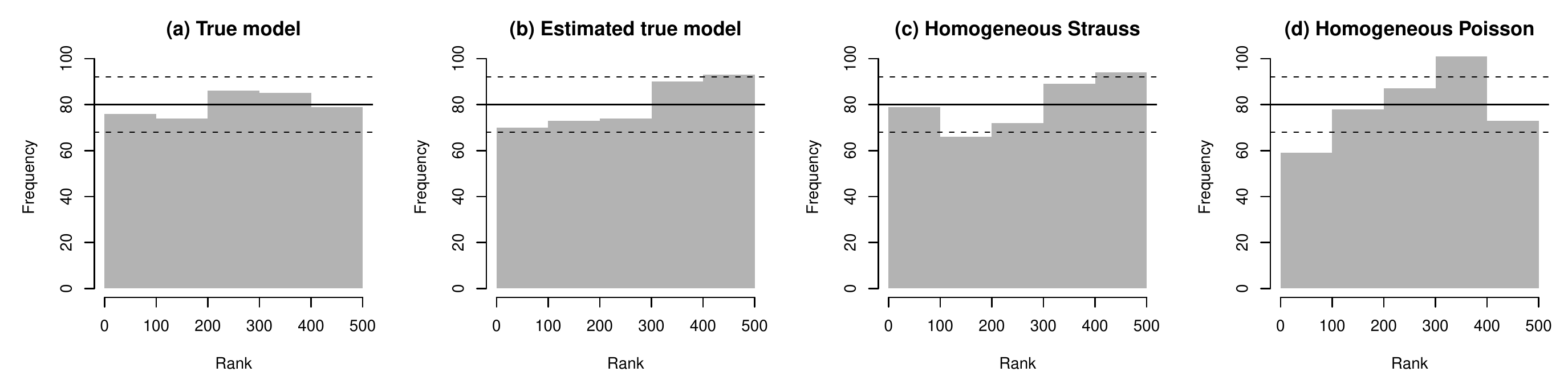}
\caption{Verification rank histograms for the inhomogeneous Strauss pattern in Figure~\ref{fig:Strauss} aggregated over $400$ pixels under (a) the true model, (b) the estimated correctly specified model, (c) a homogeneous Strauss model and (d) a homogeneous Poisson model. The pointwise 90\% confidence intervals (dashed lines) are based on a bootstrap of 500 realizations from the true model.}
\label{fig:Strauss_hist}
\end{figure}

The spatial maps of the ranks are shown in Figure~\ref{fig:Strauss}.  As in the previous example, we notice an inhomogeneity in the maps for models with incorrect first order structure, the ranks are too high is areas with high intensity and too low in areas with low intensity.  Here, we also see a clear divergence from uniformity in the rank histograms, see Figure~\ref{fig:Strauss_hist}.  The rank histogram for the estimated yet correctly specified model is slightly biased towards higher ranks.  The simulated data sets under this model have on average $259$ data points, with a range equal to $[228, 288]$, which could explain this effect.  

The rank histogram for the homogeneous Strauss model exhibits a similar effect.  Here, the model seems even more biased with the size of the simulated patterns ranging from $206$ to $276$.  The histogram is furthermore slightly $\cup$-shaped indicating a small amount of underdispersion.  The Poisson model, on the other hand,  has an overdisperse  $\cap$-shaped histogram indicating that the Poisson patterns have simultaneously both more gaps and more clusters than the original pattern.  This follows from the Poisson patterns being more clustered than the repulsive Strauss pattern while having, on average, the same number of points as the original pattern.  

\subsection{Inhomogeneous Poisson process}

\begin{figure}[t]
\centering
\includegraphics[width=13cm]{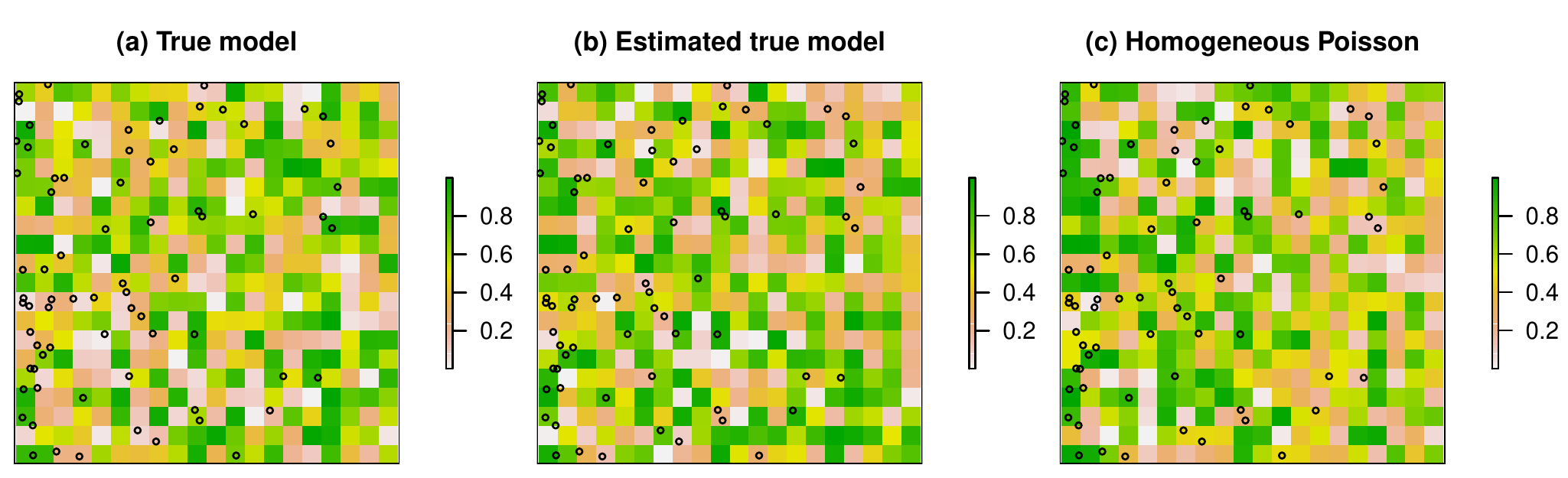}
\caption{Spatial maps of the PIT values for an inhomogeneous Poisson process on the unit square under three different models; (a) Poisson model with the true intensity $\lambda(u)=300 \exp(-3u_1)$, (b) Poisson model with estimated intensity $\hat{\lambda}(u)=223 \exp(-2.89u_1)$, (c) Poisson model with estimated intensity $\hat{\lambda} = 73$.  The point pattern is indicated with small circles.}
\label{fig:Po300}
\end{figure}

\begin{figure}[!hbpt]
\centering
\includegraphics[width=13cm]{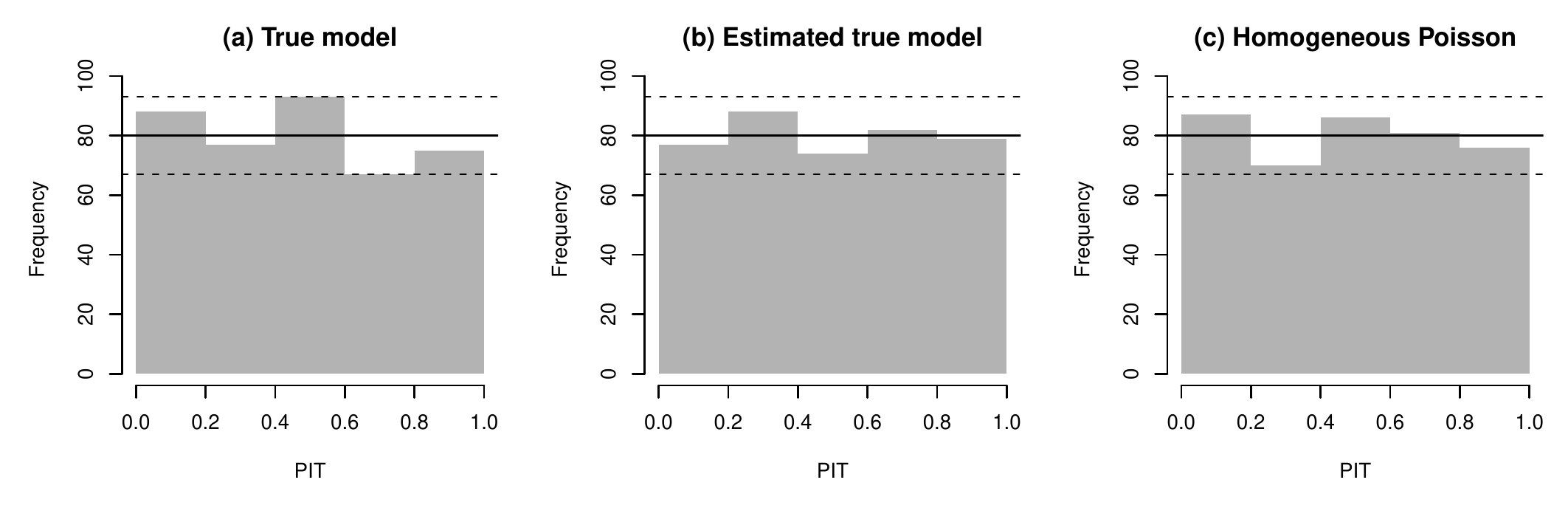}
\caption{Histograms of the PIT values for an inhomogeneous Poisson process on the unit square based on 400 equally sized pixels under three different models; (a) Poisson model with the true intensity $\lambda(u)=300 \exp(-3u_1)$, (b) Poisson model with estimated intensity $\hat{\lambda}(u)=223 \exp(-2.89u_1)$, (c) Poisson model with estimated intensity $\hat{\lambda} = 73$.  The dashes lines indicate the $90\%$ point wise confidence intervals under a null hypothesis of i.i.d. standard uniform values.}
\label{fig:Po300_hist}
\end{figure}

An inhomogeneous Poisson process  $\varphi = \{ \xi^i\}_{i=1}^n$ on a finite observation window $W \subset \real^2$ has density 
\begin{equation}\label{eq:Poisson}
p_{\theta}(\varphi) = \exp \Big( |W| - \int_W \lambda_{\theta}(u)\dd u  \Big) \prod_{i=1}^n \lambda_{\theta}(\xi^i)
\end{equation}
with respect to the unit rate Poisson process, where $\lambda_{\theta}$ is an inhomogeneous intensity function and $\theta$ denotes the parameters of the model.  Here, we consider the window  $W = [0,1] \times [0,1]$ and define $\lambda_{\theta}(u) = \theta_1 \exp(- \theta_2 u_1)$ with $\theta_1 = 300$ and $\theta_2 = 3$. The simulated pattern has $73$ points and is shown in Figure~\ref{fig:Po300}.  Maximum likelihood estimation yields $\hat{\theta}_1 = 223$ and $\hat{\theta}_2 = 2.89$.

For the calibration assessment, we divide $W$ in $400$ equally sized pixels as shown in Figure~\ref{fig:Po300} with between $0$ and $3$ points falling in each pixel.  The expected number of points in each pixel as given in (\ref{eq:poisson parameter}) can be calculated explicitly and we may thus apply the randomized PIT in (\ref{eq:PIT}).  The spatial maps in Figure~\ref{fig:Po300} show no apparent differences between the true model and the estimated true model, while a slight skewness in the spatial distribution of the PIT values under a homogeneous model is apparent.  However, the histograms in Figure~\ref{fig:Po300_hist} show no significant differences between the correctly and the incorrectly specified models.

We may obtain an increased sensitivity in the PIT histogram by increasing the observation window $W$.  For instance, if we consider 10 independent repetitions of the point pattern in Figure~\ref{fig:Po300} or, equivalently, set $W = [0,1] \times [0,10]$,  the PIT histogram for the misspecified homogeneous model diverges significantly from uniformity.  Similar results hold if we increase the expected number of points in the pattern, say, with a true intensity $\lambda(u) = 1000 \exp(-3u_1)$ (results not shown).  

\begin{comment}
\begin{figure}[!hbpt]
\centering
\includegraphics[width=13cm]{inhom_ranks1000}
\caption{Spatial maps of the PIT values for an inhomogeneous Poisson process on $W = [0,1] \times [0,1]$ under three different models; (a) Poisson model with the true intensity $\lambda(u)=1000 \exp(-3u_1)$, (b) Poisson model with estimated intensity $\hat{\lambda}(u)=928 \exp(-2.87u_1)$, (c) Poisson model with estimated intensity $\hat{\lambda} = 305$.  The point pattern is indicated with small circles.}
\label{fig:Po1000}
\end{figure}

\begin{figure}[!hbpt]
\centering
\includegraphics[width=13cm]{inhom_histograms1000}
\caption{Histograms of the PIT values for an inhomogeneous Poisson process on $W = [0,1] \times [0,1]$ based on 400 equally sized pixels under three different models; (a) Poisson model with the true intensity $\lambda(u)=1000 \exp(-3u_1)$, (b) Poisson model with estimated intensity $\hat{\lambda}(u)=928 \exp(-2.87u_1)$, (c) Poisson model with estimated intensity $\hat{\lambda} = 305$.  The dashes lines indicate the $90\%$ point wise confidence intervals under a null hypothesis of i.i.d. standard uniform values.}
\label{fig:Po1000_hist}
\end{figure}
\end{comment}

\subsection{Homogeneous cluster process}

\begin{figure}[t]
\centering
\includegraphics[width=13cm]{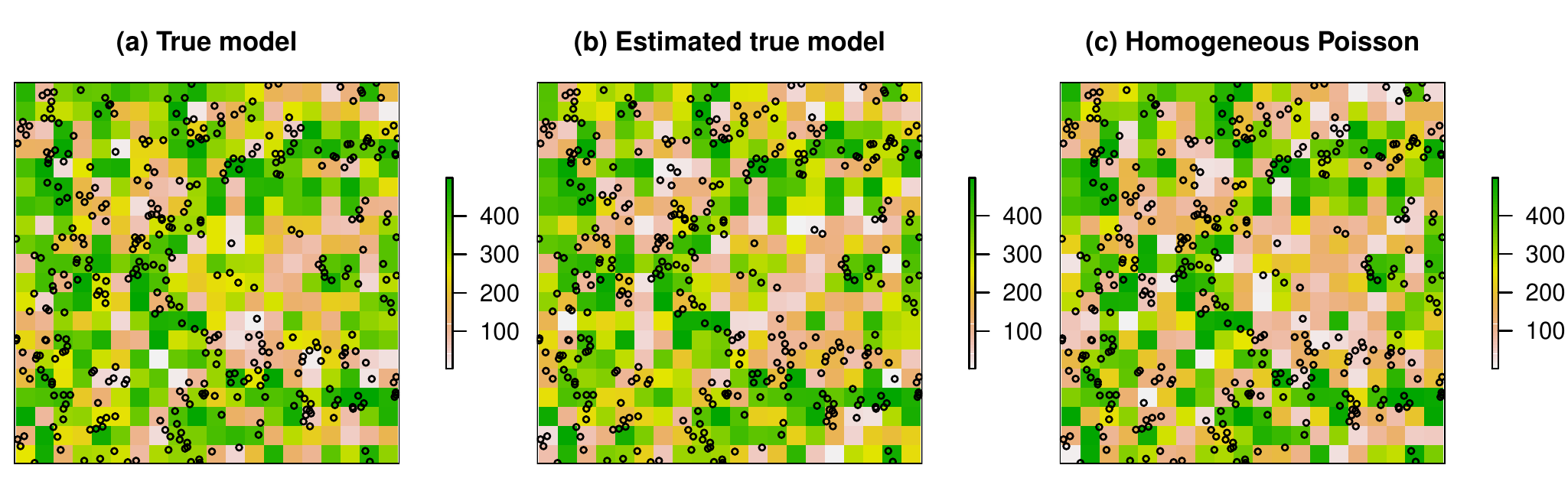}
\caption{Spatial maps of observed ranks of a pattern drawn from Geyer saturation model in (\ref{eq:Geyer}) under (a) the true model, (b) the estimated correctly specified model and (c) a homogeneous Poisson model. The ranking is based on 499 samples from each model and the ranks thus range from 1 to 500. The point pattern is indicated with small circles.}
\label{fig:Geyer}
\end{figure}

Our last example is a realization of a homogeneous Geyer saturation process \citep{Geyer1999} with density
\begin{equation}\label{eq:Geyer}
p_{\theta}(\varphi) = \mathcal{Z}(\theta) \exp \big( n(\varphi) \log \beta + s_{\theta}(\varphi) \log \gamma \big), 
\end{equation} 
where the interaction function is given by $s_{\theta}(\varphi) = \sum_i \min \big\{ \alpha, \sum_{j: j \neq i} \mathbbm{1} \{ \| \xi^i - \xi^j \| \leq r\} \big \}$ for a saturation threshold $\alpha$.  This is an example of a homogeneous process with moderately strong clustering. As in the previous example, we fix the parameters of the interaction function and set $r = 0.05$ and $\alpha = 4.5$. Then, we apply maximum pseudolikelihood to estimate the remaining parameters $\theta = (\beta, \gamma)$ which have true values $\beta = \exp(4)$ and $\gamma = \exp(0.4) \approx 1.5$. The estimation returns $\log \hat{\beta} = 4.12$ and $\hat{\gamma} = 1.46$. For comparison, we also consider the true model as well as a homogeneous Poisson model with estimated intensity $\hat{\lambda} = 376$, the number of data points in the pattern. 

\begin{figure}[t]
\centering
\includegraphics[width=13cm]{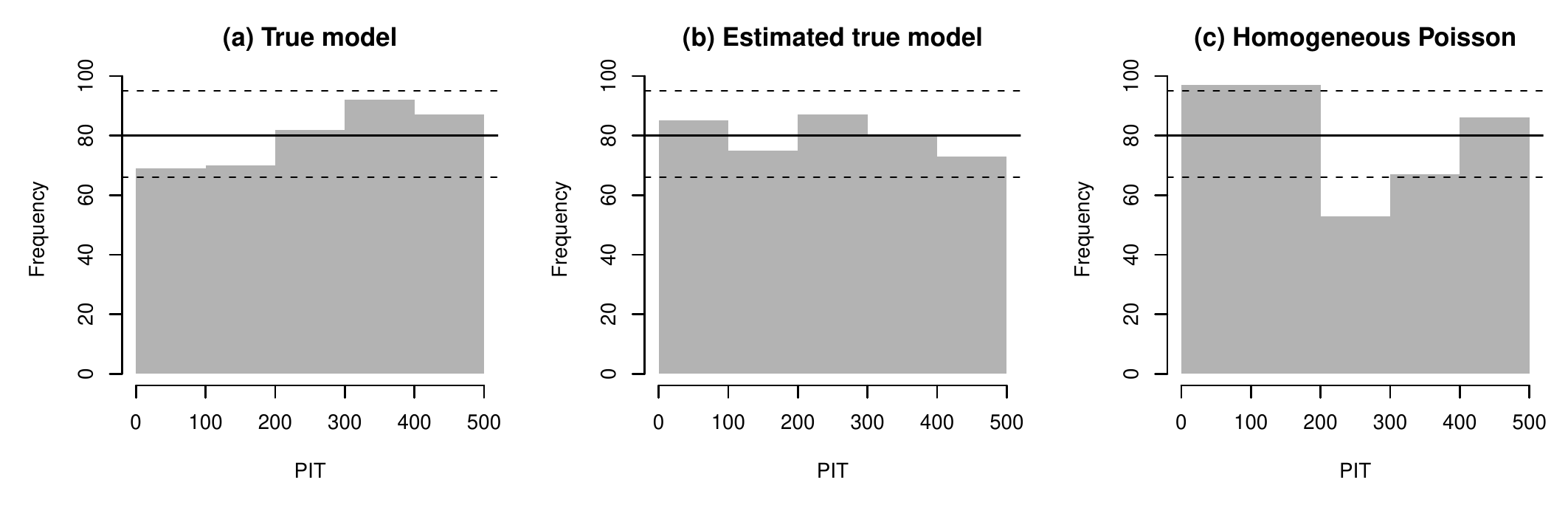}
\caption{Verification rank histograms for the point pattern in Figure~\ref{fig:Geyer} aggregated over 400 pixels on the unit square under (a) the true Geyer saturation model, (b) the estimated correctly specified model and (c) a homogeneous Poisson model.  The pointwise 90\% confidence intervals (dashed lines) are based on a bootstrap of 500 realizations from the true model.}
\label{fig:geyer_hist}
\end{figure}

As before, the observation window is the unit square divided in 400 equally sized pixels; there are now between 0 and 6 observed points in each pixel.  The spatial maps in Figure~\ref{fig:Geyer} are quite similar for all three models, while the verification rank histogram for the Poisson model in Figure~\ref{fig:geyer_hist} is $\cup$-shaped indicating that the model is underdispersive. That is, the point pattern is more clustered and there are more empty pixels than one would expect for a Poisson model with a similar number of points. 

\section{Discussion}

We propose to employ the model evaluation framework of \cite{Dawid1984}, \cite{Diebold&1998} and \cite{Gneiting&2007} to assess the statistical consistency between a point process model and the realized point pattern. In our calibration assessment, we investigate whether events that are estimated to happen with probability $p$ occur with empirical frequency $p$ in the data.  The framework is easily applicable and holds for all models from which we can obtain random draws.  The PIT or rank histogram may reveal model bias and incorrect representation of the second order interaction structure in the data while a map of the PIT values or ranks over the observation window can demonstrate deficiencies in the first order structure.   

As with other pixel-based methods, the PIT calibration results are somewhat sensitive to the pixel size.  In particular, information will be lost if the pixels are too large while very small pixels will reduce the pixel-wise distributions to binary probabilities.  However, the distribution of the PITs doesn't suffer the same skewness as the residual distribution for small pixels and low point intensity reported e.g. in \cite{Clements&2011} due to the use of the randomized PIT in (\ref{eq:PIT}).  In our examples, we have used 5 histogram bins for 400 values where the observed point patterns have between 0 and 6 points in each pixel.  This seems to give sufficiently robust results though it should be noted that with only 400 values, a theoretically uniform histogram might diverge substantially from a flat histogram, see the confidence bounds in Figure~\ref{fig:Po300_hist}.  

While we have focused on spatial point processes, it is straightforward to apply the PIT or rank calibration diagnostics to temporal or space-time processes as well.  For marked point processes, on the other hand, alternative methods are called for.  \cite{Schoenberg2003} considers thinned residuals for the space-time-magnitude distribution of earthquake occurrences and \cite{CoeurjollyLavancier2013} propose an extension of the residual framework of \cite{Baddeley&2005} that applies to stationary marked Gibbs processes. In our setting, marked point processes might be dealt with by applying the multivariate rank histogram proposed by \cite{Gneiting&2008}, where the mark distribution is considered a separate component of a multivariate distribution. 

%\begin{verbatim}
%\begin{figure}
%\centering
%\includegraphics{<figure name>}
%\caption{<Figure caption>}
%\end{figure}
%\end{verbatim}

%\ack{I would like to thank Tilmann Gneiting, Adrian Baddeley and Alex Lenkoski for sharing their thoughts and expertise.  This work which was supported by Statistics for Innovation, $sfi^2$, in Oslo.}

\section*{Acknowledgment}
I would like to thank Tilmann Gneiting, Adrian Baddeley, Alex Lenkoski and Peter Guttorp for sharing their thoughts and expertise.  This work was supported by Statistics for Innovation, $sfi^2$, in Oslo.

% In using BibteX, use wb_stat.bst
\bibliographystyle{wb_stat}
\bibliography{PointProcess}

\end{document}